\documentclass[journal]{IEEEtran}

\usepackage{amsmath,amsfonts}
\usepackage{array}
\usepackage{textcomp}
\usepackage{stfloats}
\usepackage{url}
\usepackage{graphicx}
\usepackage{cite}
\usepackage{lmodern}
\usepackage[T1]{fontenc}
\begin{document}

\title{Quality-of-Transmission Estimation in Quantum Networks: Compute, Simulate, or Learn?}

\author{
Anis~Bekri$^{1,2}$,
Amar~Abane$^{1}$,
Van~Sy~Mai$^{1}$,
Abdella~Battou$^{1}$,
and Saddek~Bensalem$^{2}$%
\thanks{$^{1}$National Institute of Standards and Technology, Gaithersburg, Maryland, USA.}%
\thanks{$^{2}$Universit\'e Grenoble Alpes, Grenoble, France.}%
\thanks{This work has been submitted to the IEEE for possible publication. Copyright may be transferred without notice, after which this version may no longer be accessible.}
}

\maketitle

\begin{abstract}
Future quantum networks will rely on control-plane decisions such as routing,
scheduling, entanglement provisioning, purification, and memory allocation. Each
decision depends on a timely and reliable estimate of the quality of the
entanglement that a candidate path can deliver, with end-to-end fidelity as a
central metric. Inspired by the optical-networking experience with
quality-of-transmission (QoT) estimation, this article asks when a quantum
controller should compute fidelity analytically, simulate it, or learn a
surrogate model. The optical analogy is useful but incomplete: optical QoT
estimation often predicts a scalar transmission margin, whereas quantum QoT
estimation must account for probabilistic entanglement generation, random memory
waiting times, decoherence, state consumption, and measurement constraints that
can make fidelity a stochastic target. We propose a compute/simulate/learn
framework for quantum QoT estimation and illustrate it with a simulation-based
proof of concept. Our study shows that analytical composition can be sufficient
in simple regimes, that simulation becomes necessary when stochastic memory
effects shape the delivered fidelity distribution, and that lightweight learned
surrogates can approximate simulator labels at much lower runtime cost for online
control. We close with open challenges for deployment, including stochastic
targets, sim-to-real transfer, data-generation cost, physics-informed learning,
and generalization beyond individual paths to full network topologies.
\end{abstract}

\begin{IEEEkeywords}
Quantum networks, entanglement distribution, fidelity estimation, quality of
transmission, machine learning, network simulation.
\end{IEEEkeywords}

\section{Introduction}

Quantum networks are expected to support applications that rely on distributed
entanglement, from secure communication to distributed sensing and networked
quantum computing~\cite{wehner_vision}. Delivering such services is not only a
physical-layer problem.
A quantum network also needs a control plane able to decide which path should be
used for an entanglement request, which elementary links should be generated
first, when entanglement swapping should be attempted, whether purification is
worth the cost, and how scarce quantum memories should be allocated~\cite{routing_survey}. Each of these decisions depends on a prior estimate of the quality that a candidate path can deliver.

The central metric in this article is the end-to-end fidelity of the delivered
entangled state, considered together with rate and waiting time. These are the
quantities in which quantum applications are expected to express their
quality-of-service (QoS) requirements~\cite{qos_quantum}. This makes
fidelity estimation the quantum counterpart of quality-of-transmission (QoT)
estimation for optical lightpaths, where QoT estimators help decide whether a candidate
connection will meet the minimum signal quality required for reliable reception. These
estimators may rely on analytical models, physical-layer tools such as
GNPy~\cite{gnpy}, monitoring data, or learned predictors for QoT estimation of
established and unestablished lightpaths~\cite{damico,ibrahimi,khan}. A similar
operational question now appears in quantum networking: before committing
resources to a path, the controller needs to estimate whether the resulting
entanglement will satisfy certain QoS requirements.

The analogy with optical QoT estimation is useful, but incomplete. In optical
networks, QoT is often summarized by a scalar margin, such as the optical
signal-to-noise ratio (OSNR), the generalized signal-to-noise ratio (GSNR), or a predicted bit error rate (BER). For a fixed lightpath, this margin is usually treated as a deterministic quantity to estimate conservatively. In quantum networks, the same
path can deliver a distribution of fidelities across requests because
entanglement generation is probabilistic, memories decohere while waiting, and
swapping or purification may succeed or fail~\cite{sequence,netsquid}. The mean of this distribution is a
useful first target for control-plane decisions, and it is the target studied in
our proof of concept. More generally, treating fidelity as a distribution opens the
door to risk-aware QoT estimation, where a controller may also use lower
quantiles or threshold-violation probabilities when allowed. This is, however, outside the scope of this paper.

There are three natural ways to obtain such an estimate. An analytical formula is fast and interpretable, but it is only reliable when its assumptions match the operating regime and when its parameters are fresh and observable. A simulator can model richer dynamics, including waiting times and memory decoherence, but it is too expensive to run inside every online routing or scheduling decision~\cite{sequence,netsquid}. A learned model can provide fast
predictions after training and can therefore serve as a surrogate for expensive
simulation, an idea now emerging in quantum-network optimization~\cite{surrogate}.

This article organizes quantum fidelity estimation around a simple operational
principle: the estimator should adapt to the regime in which the controller
operates. For simple, well-characterized paths, analytical composition provides a
fast and transparent estimate. However, when probabilistic generation, memory
waiting times, and complex decoherence patterns shape the delivered fidelity
distribution, simulation provides the right level of detail. Finally, when many
candidate paths or resource-allocation decisions must be evaluated online, a
learned model can turn simulator-generated data into a fast estimator. The
practical question is therefore the following: when should a controller compute
fidelity analytically, when should it simulate, and when should it rely on a
learned surrogate model? We present a simple framework that helps make that
decision, and we illustrate it with a quantum-network simulator-based study.

\section{Quantum QoT and the Optical Precedent}

\subsection{What ``QoT'' Means in a Quantum Network}
\begin{figure*}[!t]
\centering
\includegraphics[width=\textwidth,height=0.35\textheight,keepaspectratio]{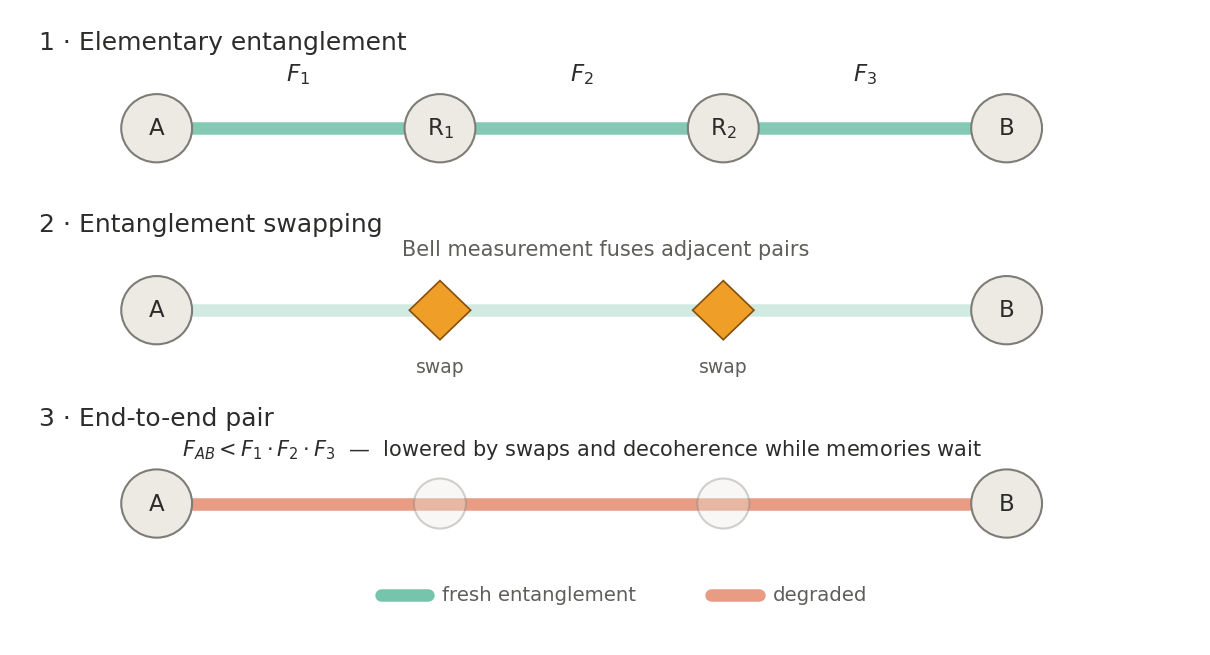}
\caption{Entanglement distribution along a repeater path. Elementary links are
generated probabilistically and then fused by Bell-state measurements at
intermediate repeaters. The final end-to-end fidelity is affected not only by
the elementary-link qualities, but also by swap imperfections and memory
decoherence accumulated while neighboring links wait.}
\label{fig:mechanism}
\end{figure*}

The QoT of a quantum path is carried chiefly by the
end-to-end fidelity. It is hard to predict for reasons with no optical analogue:
it is inherently stochastic, decoheres in memory, is consumed on use, and the
no-cloning theorem forbids measuring it without destroying the state. Entanglement
is distributed by generating short elementary links and fusing them through
entanglement swapping at intermediate repeaters (Fig.~\ref{fig:mechanism}); each swap
extends reach but compounds imperfection and lets decoherence act over the
accumulated waiting time.

For a path $\mathcal{P}_{s,d}$ of Werner links\footnote{A Werner link is an
elementary entangled channel whose shared state is modeled as a mixture of a
maximally entangled Bell state and uniform noise.} with fidelities $F_\ell$,
ideal swapping composes the end-to-end fidelity multiplicatively:
\begin{equation}
\label{eq:werner}
F_{s,d}=\frac{1}{4}\!\left(1+3\!\!\prod_{\ell\in\mathcal{P}_{s,d}}\!\!W_\ell\right),
\qquad W_\ell=\frac{4F_\ell-1}{3}.
\end{equation}
Equation~\eqref{eq:werner} is the quantum counterpart of the optical GSNR
composition: a highly efficient and common formula that the network control plane
can easily utilize.

\subsection{Three Ways to Get the Metric}
Table~\ref{tab:sources} contrasts the three sources along cost, accuracy, and the
parameter knowledge they require. Analytical formulas are fast but their accuracy is regime-dependent: exact when little decoherence occurs, biased once memories wait. Simulators are accurate across regimes because they reproduce the
underlying stochastic dynamics, but that accuracy comes at a real runtime cost, since estimating an expected fidelity means averaging
many simulation runs. Learned surrogates pay a one-time data-generation and training cost and then provide rapid predictions from features, implicitly absorbing whatever physics the formula leaves out.

\subsection{The Optical QoT Estimation Story}
Optical networks faced this same estimation problem a decade ago. A physics
engine such as GNPy~\cite{gnpy} computes the generalized signal-to-noise ratio of a lightpath
from fiber and amplifier parameters, and machine-learned QoT estimators were
layered on top to predict margins for lightpaths that had never been provisioned,
tightening design margins and recovering capacity~\cite{damico,ibrahimi,khan}.
The lesson we import is the decision discipline itself: compute where the physics
is trustworthy and cheap, and learn where estimates must be produced online for
configurations never seen at design time. In optics, an analytical model is
usually accurate once its parameters are supplied. Learning earns its place
because those parameters are impractical to probe exhaustively across a live
network. In the quantum case, as the next section argues, the deployable
formula's weakness is different in kind. It is undermined less by unknown
parameters than by a dynamical effect, namely decoherence over random waiting
times, that its functional form cannot represent at all, together with inputs
that no-cloning makes unobservable at runtime.

\begin{table*}[!t]
\caption{Sources of a Quantum QoT Estimate}
\label{tab:sources}
\centering
\small
\begin{tabular}{lcccc}
\hline
Source & Offline cost & Online cost & Accuracy & Parameters needed \\
\hline
Analytical & None & Low & Regime-dependent & Exact, fresh \\
Simulator & None & High & High & Exact, known \\
Learned surrogate & High & Low & Training-dependent & Implicit, from data \\
\hline
\multicolumn{5}{l}{\footnotesize Offline cost for the surrogate covers data generation and training.} \\
\end{tabular}
\end{table*}

\section{Learn, Compute, or Simulate? A Decision Grid}
When the control plane needs the fidelity of a candidate request, ``use machine
learning'' is often treated as a single decision. It is in fact two decisions and one workflow. The first decision is whether a learned predictor should replace the cheap analytical \emph{formula} and the second is whether it should replace the accurate but slow \emph{simulator}. If learning is selected, the question then becomes operational: simulator outputs must be collected offline, converted into a QoT dataset, and used to train a surrogate that can be queried online by the control plane. We separate these choices into two axes and fold them into the decision-aware workflow of Fig.~\ref{fig:grid}.

\subsection{Axis A: Learn vs. Compute: When the Formula Fails}
The multiplicative composition described in~\eqref{eq:werner} is compelling
because it condenses an entire network path into a straightforward product of
per-link values. In this context, a learned model becomes necessary only when
calculating that product is no longer appropriate, which can happen for two
distinct reasons below.

First, \emph{the formula ignores the dynamics that generate the state.}
Equation~\eqref{eq:werner} composes fixed link fidelities as if they were
delivered instantaneously. In practice, elementary links are generated by
probabilistic attempts, so a link's memory waits a random amount of time for its
neighbor to be ready, and it decoheres over that waiting period. The formula has
no notion of waiting time by construction, so it is accurate only when little
decoherence occurs and biased once it does. This is not because any input is
wrong, but because its functional form cannot represent a time-dependent effect
at all. The illustrative study below shows this gap directly: a calibrated
version of~\eqref{eq:werner} is close to exact in a low-decoherence regime and
increasingly biased as waiting times grow, which is the structured, learnable
residual a surrogate can absorb. Other simplifications compound the same
weakness. With an imperfect Bell-state measurement (BSM) at node $k$, of quality
$g_k$, two segments compose as
\begin{equation}
\label{eq:swap}
W_{ij}=g_k\,W_i\,W_j ,
\end{equation}
and assuming a single global $g$ already fails once BSM quality varies across
nodes. Interleaving purification would break the product further, since
purification is nonlinear and order-dependent. We do not measure these effects
here, but they point in the same direction: relaxing any of the formula's
idealizations turns a clean product into something a fixed formula cannot track,
while remaining learnable from the same features.

Second, \emph{the parameters are unobservable or stale.} A perfect formula is
still impractical if its inputs, such as the active link fidelities or the
residual coherence within quantum memories, cannot be readily obtained, a common
obstacle in quantum networking. No-cloning forbids reading the fidelity of a live entangled state without measuring, and thereby destroying, it; fidelity is
learned only by consuming pairs in tomography or verification protocols, or
inferred through proxies~\cite{routing_survey}. Worse, the relevant state
decoheres on timescales comparable to classical control round-trips, so any
measured value is partly out of date by the time a decision consumes it.
Continuous characterization of every link and memory is costly and competes with
live traffic for the same scarce quantum resources. This is the true analogue of
the optical ``analytical-is-infeasible'' case, where probing every spectral load
before deployment was impractical. A learned predictor sidesteps the problem by
mapping cheaply observable, stable features, such as topology, distances, nominal hardware specifications, and request type, onto the fidelity that the hidden instantaneous parameters would have produced, having seen ground truth once, at training time. This is where a learned estimator most clearly demonstrates its value, offering a perspective that is highly compelling: rather than ``learning beats physics,'' it is ``learning supplies what measurement cannot.''

\subsection{Axis B: Learn vs. Simulate: The Deployment Case}
The second axis is often mischaracterized. When a study indicates that a learned
model ``generalizes to unseen networks,'' the baseline it outperforms is not the
analytical formula, which generalizes for free, since one simply plugs in new
lengths and hop counts as inputs. The baseline is in fact re-running the
simulator. A detailed simulator is accurate and, for that reason, slow: a
simulation run cannot sit inside a routing loop that must answer in milliseconds,
and it may not even be configured for the topology or scale at hand. A surrogate
trained on simulator traces offers simulator-grade accuracy at formula-grade
speed. Given the right inductive bias, such as graph structure rather than
fixed-length feature vectors, it can also provide predictions for networks it
never saw during training. The argument here is therefore not that the formula is wrong, but that accuracy is needed without paying the simulator's cost at
decision time. Establishing the correct baseline is critical, as it dictates what the results must compare against: the study below evaluates the surrogate against re-simulation rather than against the product formula on this axis.

\subsection{Putting It Together}
\begin{figure*}[!t]
\centering
\includegraphics[width=\textwidth]{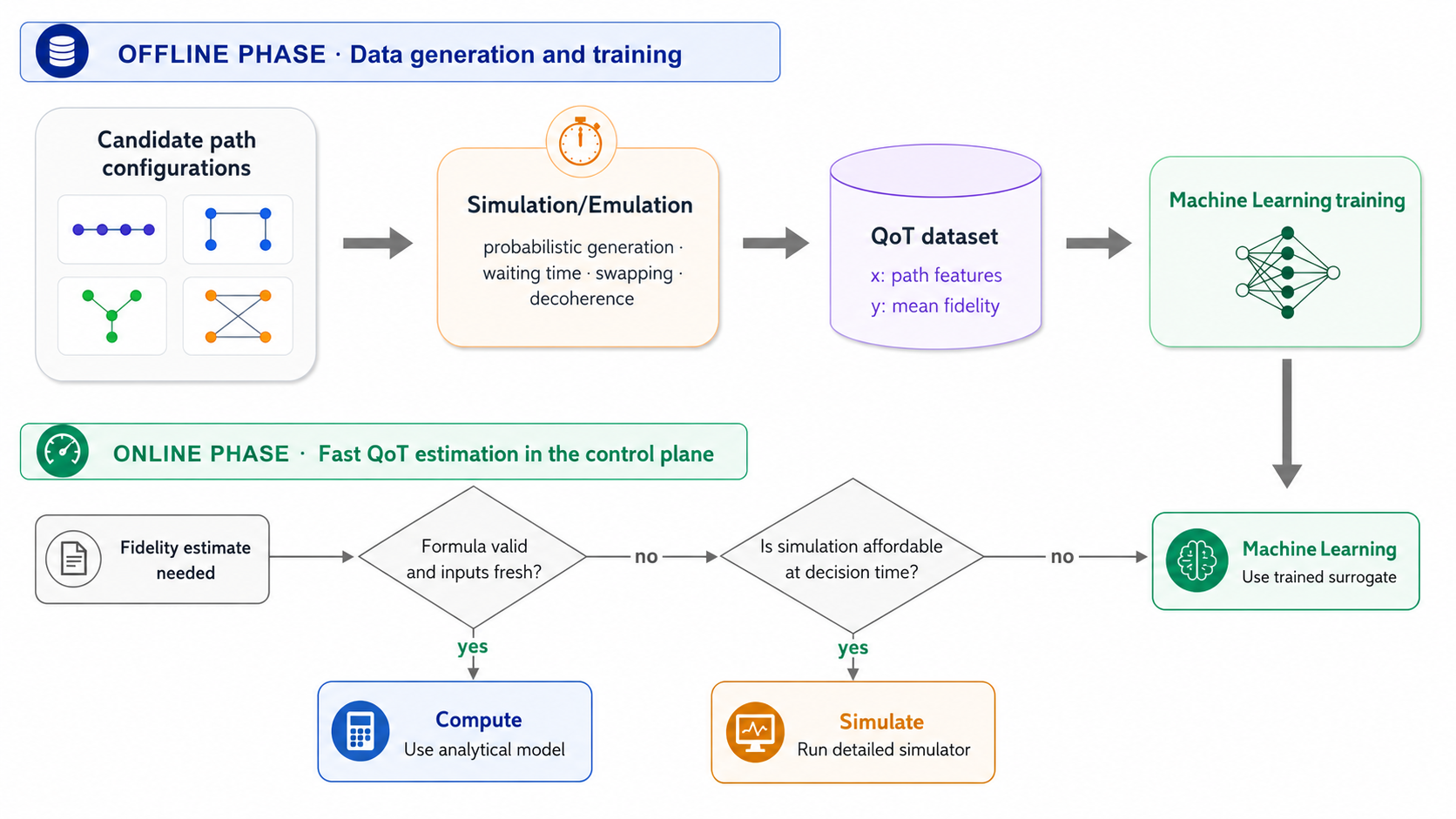}
\caption{Decision-aware machine-learning workflow for quantum QoT estimation. The upper part shows the offline training workflow: candidate path configurations are evaluated with a quantum-network simulator to produce fidelity labels, build a QoT dataset, and train a machine-learning surrogate. The lower part shows the online decision flow: a fidelity-estimation request is mapped to analytical computation when the formula is valid and its inputs are fresh, detailed simulation when runtime allows, or the trained machine-learning surrogate when repeated simulation is too costly at decision time.}
\label{fig:grid}
\end{figure*}

The two axes are embedded in the decision flow shown in the lower part of
Fig.~\ref{fig:grid}. First: is the multiplicative formula both accurate for this
path and feedable with cheap, fresh parameters? If so, use it. This is the
homogeneous, well-characterized corner where learning would only stack an
unvalidated black box on top of a trustworthy answer. If not, a higher-fidelity
source is required, which raises the second question: can detailed simulation be
run at decision time? If yes, simulate; if the answer is needed online and repeated simulation is too costly, use the trained machine-learning surrogate.

The upper part of Fig.~\ref{fig:grid} shows what this last branch requires before
deployment. Candidate path configurations are evaluated offline with a
quantum-network simulator. The resulting fidelity labels are
stored together with path and hardware features to form a QoT dataset. A
machine-learning model is then trained to learn the mapping from configuration to delivered fidelity, and the trained surrogate is deployed as a fast online
estimator. Learning, in other words, is not a shortcut around simulation; it is a way to amortize expensive simulation over many future control-plane decisions.

To conclude, machine learning should not be treated as a default for quantum QoT
estimation. It is the right tool when analytical formulas become significantly
inexact or impractical \emph{and} repeated simulation is too slow for online use. Outside of this intersection, employing it is wasted effort at best. The rest of this article shows, on simulator data, where that regime actually begins.

\section{An Illustrative Simulation-Based Study}
\label{sec:study}

We now locate the boundaries of the grid empirically. The study is built entirely on
the public SeQUeNCe simulator~\cite{sequence}, using its Bell-diagonal-state pipeline
so that each delivered pair carries an analytically tracked fidelity. Elementary
links are generated by heralded, probabilistic attempts, so a link succeeds only
after a random number of tries; its memory then waits a random time for the
neighboring link to be ready, and a depolarizing channel degrades the stored state
over that wait. Entanglement swapping fuses links into an end-to-end pair. Hardware
is set to practical values: detector efficiency $0.85$ with dark counts, memory
efficiency $0.69$, and $0.2$~dB/km fiber, while configurations sample link lengths
of $2$--$40$~km, nominal link fidelities of $0.80$--$0.99$, and coherence times
log-uniformly from $0.2$ to $5$~s, and a native memory cutoff discards pairs that waited beyond half the coherence time, rather
than delivering a heavily degraded state. This cutoff is what a real controller
would apply, and it also keeps the delivered-fidelity distribution unimodal, so
that its mean is a meaningful prediction target.

The prediction task is deliberately simple. From features a control plane can
read cheaply and keep fresh, namely hop count, per-link nominal fidelities,
distances, and a coherence-derived decoherence severity, we predict the mean
delivered fidelity of a request. Two points frame every number below. The target
is the fidelity \emph{of the simulator}, not of real hardware; and the model is
given only these observable, physical features, never the analytical estimate
itself, so that outperforming the formula is a genuine comparison rather than a
correction applied on top of it. Each label is a mean over three stochastic runs,
a stable Monte-Carlo estimate of the expected fidelity.

\subsection{The Simple Regime: the Formula Suffices}
We first characterize the low-decoherence corner: two- and three-hop paths with
links of $4$--$6$~km, nominal link fidelities of $0.92$--$0.96$, and coherence
times of $50$--$80$~s, so that memories wait a negligible fraction of their
coherence time before a pair is delivered. The simulator is then effectively
deterministic: repeated runs of the same configuration agree to within $10^{-4}$
in fidelity. The \emph{uncalibrated} formula~\eqref{eq:werner} reproduces the simulated fidelity with a mean-squared error of $2\times10^{-7}$, overestimating it by about $5\times10^{-4}$ on every configuration. This residual overestimation reflects the hardware imperfections and the small amount of decoherence that the ideal formula does not model, but it is three orders of magnitude smaller than any fidelity threshold a controller would act on.

\subsection{The Complex Regime: Fidelity Becomes a Distribution}
Lengthen the links and shorten the coherence time and the picture changes
qualitatively. The delivered fidelity is no longer a number but a distribution: pairs that assembled quickly arrive fresh, pairs that waited arrive degraded, and across a representative three-hop path the per-pair fidelity spreads with a standard deviation above $0.1$. The static formula, blind to waiting time, sits well above the truth, predicting roughly $0.86$ where the simulator's expected fidelity is about $0.64$. This gap is not measurement noise; it is the decoherence the formula cannot represent. Recovering the expectation from the simulator is possible but costly: the error of an $n$-run estimate falls only as $1/\sqrt{n}$, so halving it demands four times the runs. Accuracy has a price here, and that price is what a learned surrogate amortizes.

\subsection{Learning the Expected Fidelity}
\begin{figure*}[!t]
\centering
\includegraphics[width=\linewidth]{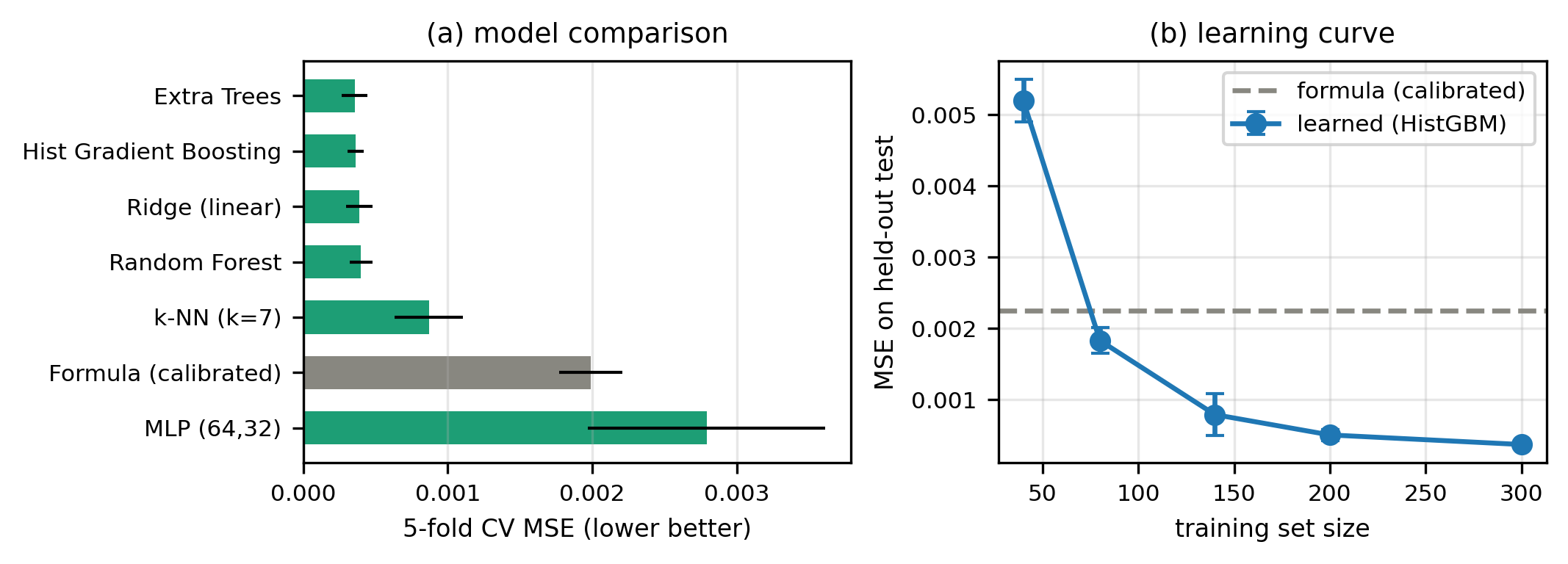}
\caption{Learning the expected fidelity. a) Five-fold cross-validated error of six
regressors against the per-fold-calibrated formula; tree ensembles cut the error
$\sim$5--6$\times$. b) Learning curve: the surrogate overtakes the formula after
$\sim$80 configurations and keeps improving.}
\label{fig:models}
\end{figure*}

On a dataset of 390 distinct configurations spanning two to four hops, we trained
a range of regressors and compared them, under five-fold cross-validation,
against the analytical formula calibrated per fold. A single effective per-hop
quality factor $g$, applied as $g^{h}$ to the Werner product of an $h$-hop path,
is fit by least squares on each training fold, so that the formula receives the
same data advantage as the learned models (Fig.~\ref{fig:models}). Tree
ensembles, namely random forests~\cite{breiman_rf}, extremely randomized
trees~\cite{geurts_extratrees}, and gradient boosting, as implemented in
scikit-learn~\cite{sklearn}, predict the expected fidelity with a mean-squared
error roughly five to six times lower than the calibrated formula, and about
$2.5\times$ lower mean absolute error. The conclusion does not hinge on model
class: a plain linear regressor performs almost as well as the tree ensembles,
which tells us that the mapping from features to fidelity is close to linear once
the right features are supplied, so a compact, interpretable model already
captures it. A neural network, the model with the most capacity, does worst of
all. It is the only learned model that fails to beat the formula, and its error
varies the most across folds. This is not a tuning problem. When the target is
simple and the dataset small, a flexible network has no rich pattern to learn and
ends up fitting noise. The ordering is itself the message: capacity beyond what
the modest, path-local target requires buys nothing, which is precisely why a
heavier structural model is unnecessary in this regime.

A learning curve (Fig.~\ref{fig:models}) shows the surrogate overtaking the
formula after roughly eighty labeled configurations and continuing to improve
without early saturation, which both justifies the dataset size and confirms that
the model is fitting signal rather than the small label noise. The operational
payoff is the point of the exercise: once trained, the surrogate returns a
prediction in well under a millisecond, against seconds for a simulator run, so
it can sit inside a control loop where re-simulation cannot.

\subsection{Generalizing to Unseen Path Regimes}
\begin{figure}[!t]
\centering
\includegraphics[width=\linewidth]{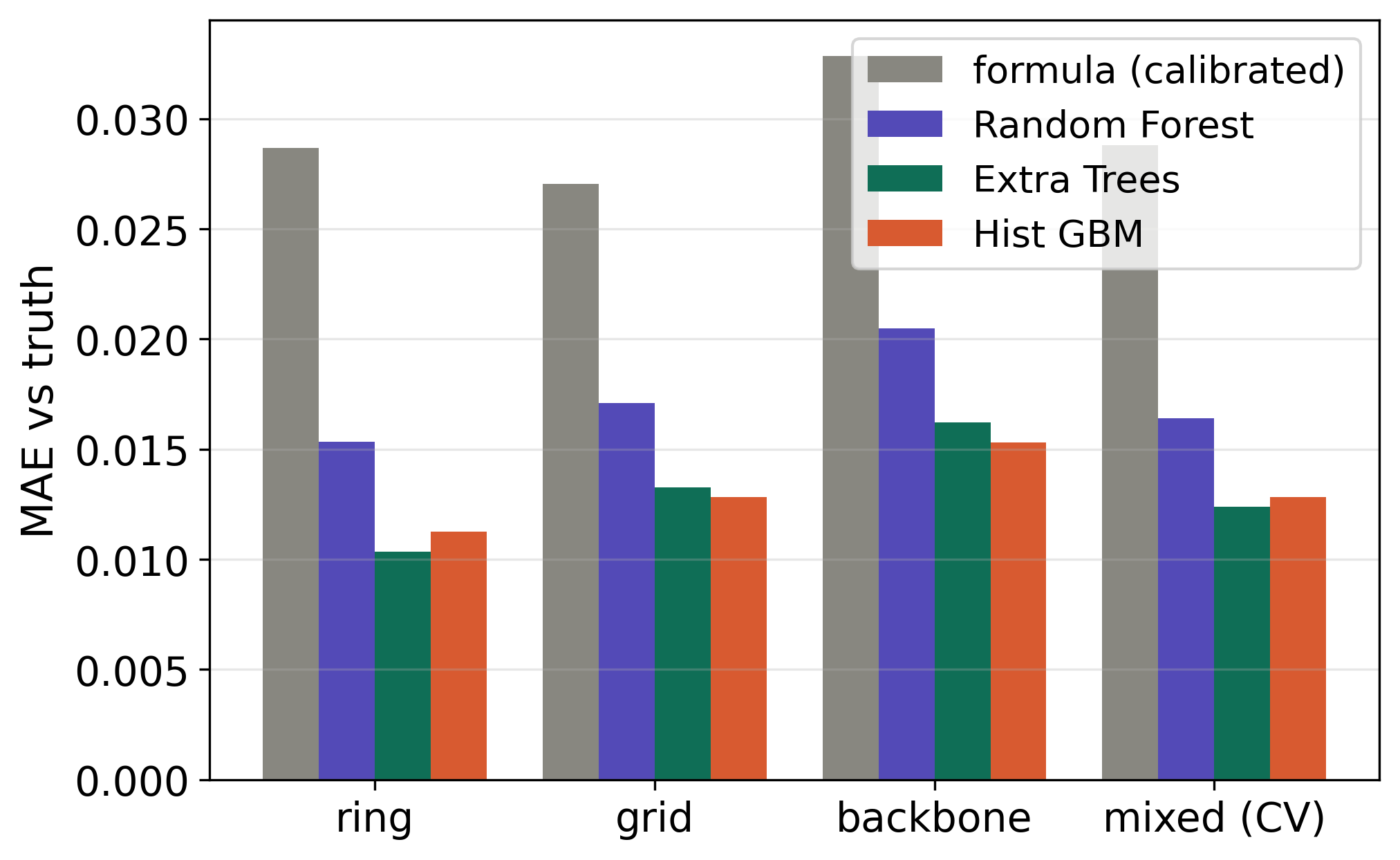}
\caption{Generalization to unseen path regimes. Leave-one-family-out error over 270
paths (90 per family; train on two of ring/grid/backbone, test on the third) beside
pooled-interpolation error;
every model beats the formula on each unseen family, and extrapolation is close to
interpolation.}
\label{fig:topo}
\end{figure}

The deployment question is not whether a surrogate fits configurations like those it
trained on, but whether it transfers to networks it has never seen, and the baseline
it must beat there is not the formula but re-running the simulator. We probe this with
three families of paths whose link-length regimes are deliberately distinct: a
short-link ``grid'' family, an intermediate ``ring'' family, and a long-haul
``backbone'' family. We train on two families and test on the unseen third
(Fig.~\ref{fig:topo}). Every model beats the calibrated formula on every held-out family, by roughly a factor of two in mean absolute error, and leave-one-family-out extrapolation is nearly as accurate as ordinary interpolation. This is evidence that the learned mapping is a genuine, transferable function of the path features rather than a memorized lookup. More precisely, these are unseen \emph{path and link regimes}, not unseen meshed topologies. In this regime the fidelity is a path-local quantity and the graph structure carries no additional signal; learning over true network structure, where a graph neural network would earn its inductive bias, requires a setting in which a path's fidelity depends on the rest of the network, which we revisit among the open challenges.

To summarize, the above study establishes a clear boundary. In the simple regime the formula is accurate enough that a surrogate is unnecessary. In the complex regime the
formula is biased due to its inability to account for decoherence. While the
simulator remains accurate, it operates too slowly for a real-time decision loop.
A learned estimator successfully bridges this gap, achieving simulator-grade
accuracy at formula-grade speed. Notably, it demonstrates this capability both on trained configurations and on unseen path regimes. That intersection, not machine learning by default, is where learning belongs.

\section{From Estimation to Action}

A fidelity estimate is only useful as it feeds a decision, and
several control-plane functions already depend on one: fidelity-aware routing
and purification schemes assume such an estimate is available at path-selection
time~\cite{li_fidelity_routing}. Path and swap-order
selection need a ranking of candidate routes before any resource is
committed. Deciding whether to purify a pair, and how many rounds, trades
delivered fidelity against rate and requires an estimate of what the pair
would otherwise arrive with. Memory allocation, that is,
how many attempts and how much waiting time to budget for a request, is itself an important factor in the fidelity a path will deliver. Each of these currently
leans on the same two options this article has been contrasting: a fast but
regime-blind formula, or infrequent and quickly stale monitoring.

A learned estimator is fast enough to rank many candidate paths or swap orders
online, which a simulator cannot be. In the provisioning language optical networks already use, this narrows the fidelity margin a controller must reserve against uncertainty, which translates directly into fewer wasted purification rounds and less over-allocated memory.

Finally, while this study focuses on establishing the foundational premise, that
is, delivering an accurate, low-latency fidelity prediction suitable for a control loop, it sets the stage for further development. Future research can extend these findings to integrate routing and scheduling policies, addressing complex challenges such as contention and joint optimization across concurrent requests.

\section{Open Challenges}
Several questions decide whether learned estimators reach deployment:

\begin{itemize}
\item \emph{Sim-to-real.} There is no corpus of real quantum-network telemetry.
How do we validate and transfer to hardware?
\item \emph{Data-generation cost.} Active learning and design of experiments
confront the surrogate paradox: we simulate in order to learn not to simulate.
\item \emph{Stochastic targets.} Fidelity is a random variable; predicting
distributions or quantiles enables risk-aware provisioning.
\item \emph{Physics-informed models.} Encoding the Werner and decay structure and
learning only the residual buys sample efficiency and interpretability.
\item \emph{Structural generalization.} Graph neural networks are the right tool
for axis B, and for joint fidelity, rate, and latency targets.
\end{itemize}

\section{Conclusion}
This article studied when machine learning is useful for estimating end-to-end
fidelity in quantum networks. We have argued that the choice among computing,
simulating, and learning is a structured decision, not a default. Learning should
replace computation only when the deployable formula is biased or unfeedable, and
it should replace simulation only when accuracy is needed faster, or at a scale,
that repeated simulation cannot provide. Learning is most useful at the
intersection of those two conditions.

Our simulation-based study located that intersection concretely. In a
well-provisioned regime, the analytical formula is accurate enough that a
surrogate is unnecessary. Once decoherence over random waiting times sets in, the
formula is biased by a dynamical effect it cannot represent. While the simulator
remains accurate, it operates too slowly for a real-time decision loop. A learned
estimator bridges this gap, achieving simulator-grade accuracy at formula-grade
speed from physical features alone, on unseen path regimes as well as familiar
ones.

The optical QoT experience supplies the roadmap, but the quantum twists, namely
decoherence, no-cloning, and consumption, redefine which regime dominates and,
with it, when learning is truly the right tool.

\section*{Disclaimer}
Any mention of commercial products or reference to commercial organizations is
for information only; it does not imply recommendation or endorsement by the
National Institute of Standards and Technology, nor does it imply that the
products mentioned are necessarily the best available for the purpose.

\bibliographystyle{IEEEtran}
\bibliography{References}

\end{document}